\documentclass[12pt]{article}
\usepackage{epsfig}
\usepackage{amsfonts}
\begin{document}
\begin{titlepage}
\begin{center}

{\Large On the small-scale statistics of Lagrangian turbulence}

\vspace{2.cm} {\bf Christian Beck}\footnote{
permanent address: School of Mathematical Sciences, Queen Mary,
University of London, Mile End Road, London E1 4NS.}

\vspace{0.5cm}

Isaac Newton Institute for Mathematical Sciences,
University of Cambridge, 20 Clarkson Road, Cambridge CB3 0EH, UK

\vspace{5cm}

\end{center}

\abstract{We provide evidence that the small-scale statistics of
the acceleration of a test particle in high-Reynolds number Lagrangian
turbulence is correctly  described by Tsallis statistics with entropic index
$q=\frac{3}{2}$. We present theoretical arguments why Tsallis statistics can
naturally arise in Lagrangian turbulence and why at the smallest scales
$q=\frac{3}{2}$ is relevant. A generalized Heisenberg-Yaglom formula is
derived from the nonextensive model.}

\vspace{1.3cm}

\end{titlepage}

Recently, methods from nonextensive statistical mechanics \cite{tsa1, tsa2,
tsa3} have been successfully applied to fully developed turbulent flows
\cite{hydro, ari1, ari2, ramos}. As a driven noneqilibrium system a turbulent
flow cannot extremize the Boltzmann-Gibbs entropy --- it is obvious that
ordinary statistical mechanics fails to provide a correct description of
turbulence. But there is experimental and theoretical evidence that the
statistics of velocity differences is well described if one assumes that the
flow extremizes the `second best' information measures available. These are
the Tsallis entropies \cite{tsa1} defined by
\begin{equation}
S_q= \frac{1}{q-1} \left( 1- \sum_i p_i^q \right).
\end{equation}
$q\not=1$ is the entropic index and $p_i$ are the probabilities associated
with the microstates of the physical system. The Tsallis entropies,
closely related to (but different from) the R\'{e}nyi information
measures \cite{renyi,BS}, are convex
and distinguished by generalized Khinchin axioms \cite{abe}. For $q\to 1$ they
reduce to the ordinary Boltzmann-Gibbs entropy. Their importance has been
demonstrated in numerous recent papers (see \cite{web} for a detailed
listing). 

Whereas previous papers on turbulence and Tsallis statistics mainly dealt with
the inertial range \cite{hydro, ari2}, in this Letter we will concentrate on
the small-scale characteristics of fully developed turbulent flows. Here one
is still far away from a complete theory, though many empirical facts of the
small-scale statistics are well known and have been experimentally verified
(see e.g. \cite{sreeni} for a review). Most turbulence measurements have been
conducted in the Eulerian frame, {\it i.e.}, by measuring the spatial
fluctuations of the velocities using the Taylor hypothesis. 
Recently, experimental progress has been made in 
investigating the Lagrangian properties of fully developed turbulence, by
tracking test particles that are advected by the flow \cite{bod1,bod2}. 
Examples of measured histograms of
the acceleration of a test particle as measured
by Bodenschatz et al. are shown in Fig.~1. The distribution has
been rescaled to variance 1. The acceleration has been extracted by parabolic fits
over $0.75 \tau_\eta$, where 
$\tau_\eta = (\nu/\epsilon)^{\frac{1}{2}}$ is the
Kolmogorov time, $\nu$ is the kinematic viscosity and $\epsilon$ 
is the average
energy
dissipation rate. 
The figure also shows as a solid line the normalized probability
density function
\begin{equation}
p(x)=\frac{2}{\pi} (1+2x^2+x^4)^{-1}= \frac{2}{\pi} (1+x^2)^{-2}, \label{cau}
\end{equation}
which has variance 1. 
Apparently, the experimental data and the above distribution function agree
quite well. We will now present theoretical arguments why this type of
distribution is relevant.

The acceleration $a$
of the Lagrangian test
particle is a strongly fluctuating random variable. It can be regarded as
a velocity difference on a small time scale $\tau$, i.e.
$a \approx (v(t)-v(t+\tau))/\tau :=u/\tau$,
where
$\tau:=\kappa
\tau_\eta$ 
is of the
order of the Kolmogorov time.
It is not clear whether the limit
$\tau \to 0$ exists in a mathematically rigorous way. For example,
already for an ordinary Brownian particle described by the Ornstein Uhlenbeck process
\cite{vKa}, the velocity exists but the acceleration is singular.
Hence it seems to make sense to statistically describe the movement
of the Lagrangian test particle 
using a small effective finite time scale $\tau$.
One can now develop the formalism of nonextensive statistical
mechanics for the fluctuating temporal changes in velocity.
Following the ideas of
\cite{hydro} we define formal energy levels $E(u)$ by the kinetic energy
of the velocity differences
\begin{equation}
E(u):= \frac{1}{2} a^2 \tau^2 \approx \frac{1}{2} (v(t)-v(t+\tau))^2=
\frac{1}{2}u^2.
\end{equation}
Moreover, a formal temperature $\beta_0^{-1}$ is introduced as
\begin{equation}
\beta_0^{-1}:=\epsilon \tau=
\epsilon\kappa \tau_\eta=\kappa \epsilon^{1/2} \nu^{1/2} \label{beta}
\end{equation}
The multiplication with
a time scale $\tau$ is necessary for dimensionality reasons,
since $\epsilon$ has dimension $length^2/time^3$.
Extremizing the Tsallis entropies one obtains the following
generalized version of a canonical distribution \cite{hydro}
\begin{equation}
p(u)= \frac{1}{Z_q} \left (1+\frac{1}{2}(q-1)\beta_0 u^2
\right)^{-\frac{1}{1-q}}.\label{tsa}
\end{equation}
For
$q\to 1$ the above probability
density reduces to the ordinary Boltzmann factor $p(u)\sim
e^{-\frac{1}{2}\beta_0u^2}$. The value of the entropic index
$q>1$ depends on the Reynolds number and the scale (see \cite{BLS}
for precision measurements in Eulerian turbulence).
The normalization
constant $Z_q$ is given by
\begin{equation}
Z_q= \left( \frac{2\pi}{(q-1)\beta_0} \right)^{1/2} \frac{\Gamma \left(
\frac{1}{q-1}-\frac{1}{2} \right)}{\Gamma \left( \frac{1}{q-1} \right)}.
\end{equation}
For $q=\frac{3}{2}$ and if rescaled to variance 1, the distribution
(\ref{tsa})
is identical with the
distribution (\ref{cau}), which apparently is in very good agreement with the
experimental data. 

All moments of the generalized canonical distribution (\ref{tsa}) 
can be evaluated analytically.
In particular, one obtains for the second moment
\begin{equation}
\langle u^2 \rangle =\int_{-\infty}^{+\infty}
u^2 p(u)du =\frac{1}{\beta_0} \frac{2}{5-3q}
\end{equation}
This yields for the second moment of the acceleration $a=u/\tau$
\begin{equation}
\langle a^2 \rangle =\frac{1}{\beta_0 \tau^2} \frac{2}{5-3q} =
\epsilon^{3/2}
\nu^{-1/2} \frac{1}{\kappa} \frac{2}{5-3q}
\end{equation}
Thus we obtain from the nonextensive
model the Heisenberg-Yaglom relation $\langle a^2 \rangle
=a_0\epsilon^{3/2}\nu^{-1/2}$, identifying the constant $a_0$ with
\begin{equation}
a_0= \frac{1}{\kappa} \frac{2}{5-3q} \label{a0}.
\end{equation}
Gaussian statistics ($q=1$) and  
$\kappa \approx 1$ would imply $a_0\approx 1$. On the other hand, the true turbulent
small scale statistics
is much better described by $q=3/2$, which yields $a_0\approx 4$, in agreement
with direct numerical simulations for large
Reynolds numbers \cite{vedula}. The precise numerical value of $a_0$
also depends on the ratio $\kappa=\tau /\tau_\eta$ which enters into
the formal thermodynamic description via
eq.~(\ref{beta}).
Bodenschatz et al. \cite{bod2} measure distributions with
$q\approx 1.5$ and $a_0\approx 5.3$ for
large $R_\lambda$, which corresponds to $\kappa \approx 0.75$. 
Generally, a measured Reynolds number dependence of $a_0$ (as presented
in \cite{bod2})
can be translated into a measurement of the entropic
index $q$ in Lagrangian turbulence, by solving eq.~(\ref{a0}) for $q$.

Let us now argue on theoretical grounds a) why Tsallis statistics can
naturally arise
in Lagrangian turbulent flows
and b) why the entropic index is $q=\frac{3}{2}$ at
the smallest scales.

Generally, Tsallis statistics with $q>1$ can arise from ordinary statistical
mechanics (with ordinary Boltzmann factors $e^{-\beta E(u)}$) if one
assumes that the formal temperature $\beta^{-1}$ is locally fluctuating (see
\cite{wilk} for similar ideas). In our application to a turbulent flow,
$\beta^{-1}$ is identified with the product $\epsilon \tau$ of local
energy dissipation and the typical
time scale $\tau$ during which energy is
transferred. Both quantities can fluctuate.
$\beta^{-1}$ is a formal variance parameter describing
the fluctuating environment of the Lagrangian test particle, measured
relative to the movement of the particle. Using the integral represention of
the gamma function
\begin{equation}
\Gamma (z)= \int_0^\infty e^{-t}t^{z-1}dt
\end{equation}
and substituting
\begin{eqnarray}
t&=&\beta\left( E (u)+\frac{1}{(q-1)\beta_0}\right)\\ z&=&\frac{1}{q-1}
\end{eqnarray}
one may write
\begin{equation}
(1+(q-1)\beta_0 E (u))^{-\frac{1}{q-1}}= \int_0^\infty e^{-\beta
E (u)} f(\beta ) d\beta \label{marl}
\end{equation}
with
\begin{equation}
f (\beta) = \frac{1}{\Gamma \left( \frac{1}{q-1} \right)} \left\{
\frac{1}{(q-1)\beta_0}\right\}^{\frac{1}{q-1}} \beta^{\frac{1}{q-1}-1}
\exp\left\{-\frac{\beta}{(q-1)\beta_0} \right\} \label{fluc}
\end{equation}
being the $\chi^2$ distribution. The physical interpretation of
eq.~(\ref{marl}) is that due to fluctuations of $\beta$ with
probability density $f(\beta)$ the Boltzmann factor
$e^{-\beta E (u)}$ of ordinary statistical mechanics has to be replaced
by the generalized Boltzmann factor $(1+(q-1)\beta_0 E(u)
)^{-\frac{1}{q-1}}$ of nonextensive statistical mechanics. The Tsallis
distribution with fixed variance parameter $\beta_0$ effectively arises by
integrating over all possible fluctuating variance parameters $\beta$.
This illustrates why the nonextensive formalism can be relevant to
non-equilibrium systems 
(formally described by a fluctuating $\beta$) if there is
a quasi-stationary state in
probability space.

The $\chi^2$ distribution is well known to occur in many very common
circumstances (see e.g.\ \cite{hast}). For example, if one has $n$ Gaussian
random variables $X_i$ then the sum of the squares of their deviations from
their mean $\chi^2 := \frac{1}{n} \sum_{i=1}^n (X_i-\bar X)^2$ is $\chi^2$
distributed with 
$\frac{2}{q-1} =n-1$. 
Or, if $T=\beta^{-1}$ obeys a linear Langevin
equation with a constant source term and a damping that fluctuates
stochastically, one also obtains a $\chi^2$ distribution for $\beta$
\cite{wilk}.

The average of $\beta$ is given by $\overline{\beta} =\int_0^\infty \beta
f(\beta) d\beta =\beta_0$ and the variance by $\overline{\beta^2}
-\overline{\beta}^2= (q-1) \beta_0^2$.
From this one obtains a physical interpretation of
the entropic index $q$ in terms of the variance of $\beta$, namely
\begin{equation}
q=\frac{\overline{\beta^2}}{\overline{\beta}^2}. \label{vic}
\end{equation}
If there are no fluctuations of $\beta$ at all, as in ordinary statistical
mechanics, eq.~(\ref{vic}) just reduces to $q=1$, as expected.

In a turbulent flow, the variance parameter $\beta$ surrounding
the Lagrangian test particle fluctuates,
and hence Tsallis statistics can naturally arise in this
context. Since the fluctuations of $\beta$ become smaller if the volume $r^3$
over which the energy dissipation is averaged increases, $q$ 
must be a montonously decreasing function of the scale $r$.
In
fact, 
at largest scales $r$ one observes approximately Gaussian behaviour
($q\approx 1$), in the inertial range $q=1.1....1.2$ gives good fits of the
experimental data of Eulerian turbulence \cite{hydro, BLS}, 
and Fig.~1 indicates that $q\approx 1.5$
at the smallest time scales of Lagrangian turbulence.

Let us now provide a theoretical argument why $q=\frac{3}{2}$ at the smallest
scales. 
The observation is that for large $|u|$ the Tsallis distributions (\ref{tsa})
(also called student or t- distributions in the statistics textbooks
) decay as
$|u|^{-w}$, where $w=\frac{2}{q-1}$. Hence only moments $\langle |u|^m\rangle$
with $m<w-1$ exist. If $q=\frac{3}{2}$ at the smallest scale, this means $w=4$
and hence the third moment would just cease to exist at the smallest scale. If
$q$ is precisely $\frac{3}{2}$ the third moment is logarithmically divergent,
if $q=\frac{3}{2}-0^+$ it just exists. Since generally the third moment is the
most important moment in turbulence, related to average energy dissipation, the
existence of this moment is necessary for turbulence to make sense. Since $q$
is monotonically decreasing with scale $r$, one ends up with the largest
allowed value of $q$ at the smallest possible scale. This is just
$q=\frac{3}{2}-0^+$.

There is further experimental evidence for the above conjecture on the small
scale statistics. In \cite{BLS} systematic measurements of the exponent $w(r)$
were performed for Eulerian turbulence
(for a turbulent Taylor Couette flow). The measurements were
performed at distances $r$ much larger than the Kolmogorov scale $\eta$. Over
a large range of scales $r$ the measured exponents $w(r)$ were very well
fitted by a power law of the form
\begin{equation}
w(r)=4 \left( \frac{r}{\eta} \right)^\delta
\end{equation}
with $\delta \approx 0.3$. Extrapolating this down to the Kolmogorov scale
$\eta$ one obtains $w=4$ at $r=\eta$, which again supports our hypothesis.
Although for very small $r$ deviations of this power law (pointing towards
smaller values of $q$) were observed in \cite{BLS}, these deviations can be
explained by the disturbing effects of noise, which naturally shifts the
entropic index to lower $q$ values, since Gaussian white noise implies $q=1$.

Our conjecture on the small scale statistics is also consistent with a large
amount of other experimental data. In \cite{sreeni} the Reynolds number
dependence of the third and fourth moment of $u$ at the smallest possible
scales was analysed. By averaging the data of many experiments it emerged that
the 4th moment increases roughly like $\sim R_\lambda^{1/3}$, whereas the 3rd
moment stays almost constant or increases much less rapidly with $R_\lambda$
than the 4th moment. This means that at the smallest scale the 4th moment is
expected to diverge for $R_\lambda \to \infty$ and the 3rd moment may either
just exist or may weakly diverge. All these experimentally observed features
are correctly reproduced by the Tsallis distribution (\ref{tsa}) with
$q=\frac{3}{2}$.  

\subsection*{Acknowledgement}
I am very grateful to Eberhard Bodenschatz for providing me with the
experimental data displayed in Fig.~1.



\newpage

\section*{Figure captions}

\noindent {\bf Fig.~1} Experimentally measured probability density 
of the
acceleration of a test particle in Lagrangian turbulence for
Reynolds number $R_\lambda =200, 690, 970$, respectively, and comparison with
the distribution~(\ref{cau}).

\vspace{2cm}

\epsfig{file=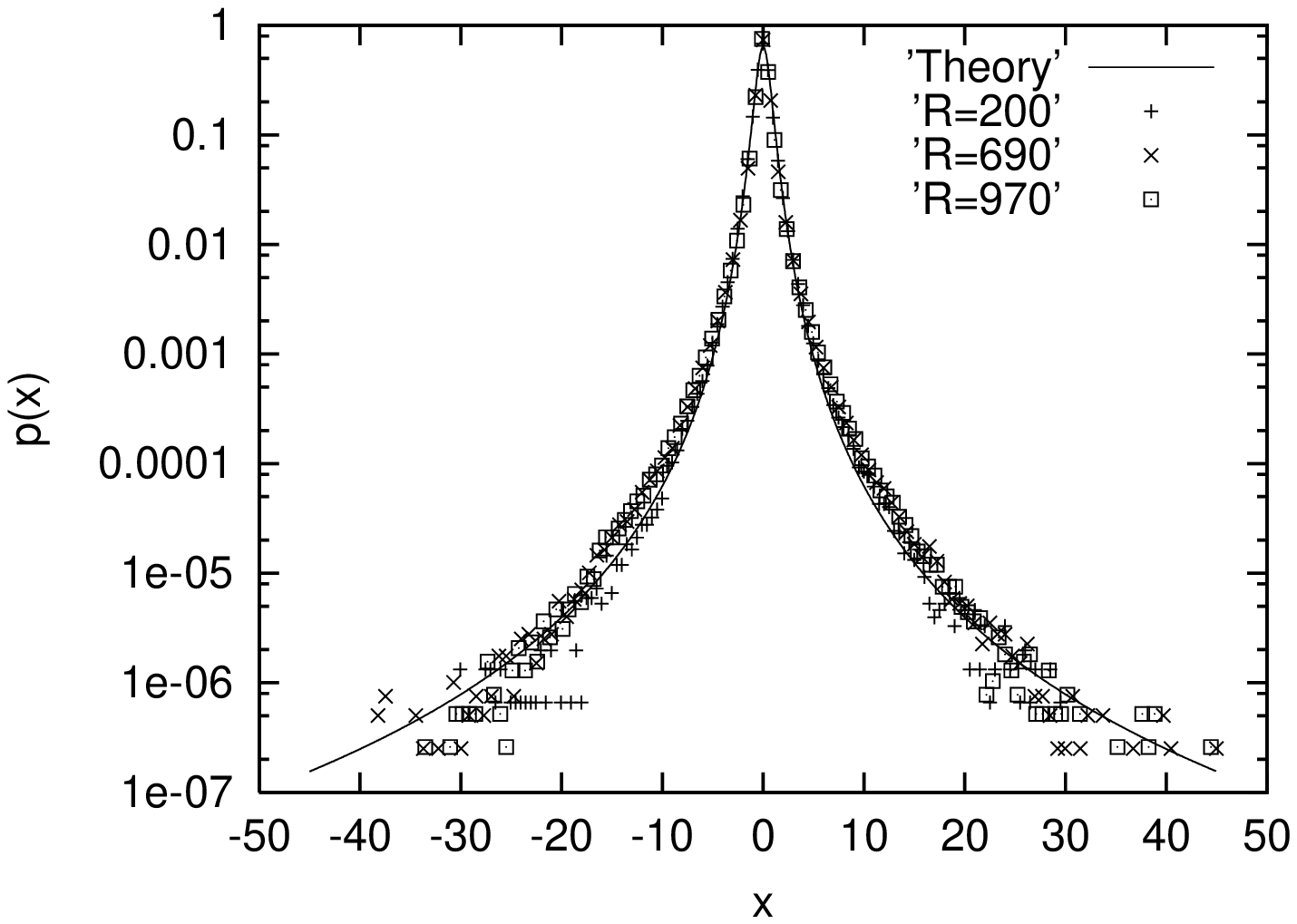, angle=-90.}

\begin{thebibliography}{99}
\bibitem{tsa1} C. Tsallis, J. Stat. Phys. {\bf 52}, 479 (1988)
\bibitem{tsa2} C. Tsallis,  R.S. Mendes and A.R. Plastino, Physica {\bf 261A},
534 (1998)
\bibitem{tsa3} C. Tsallis, Braz. J. Phys. {\bf 29}, 1 (1999)
\bibitem{hydro} C. Beck, Physica {\bf 277A}, 115 (2000)
\bibitem{ari1} T. Arimitsu and N. Arimitsu, Phys. Rev. {\bf 61E}, 3237 (2000)
\bibitem{ari2} T. Arimitsu and N. Arimitsu, J. Phys. {\bf 33A}, L235 (2000)
\bibitem{ramos} F.M. Ramos, C. Rodrigues Neto, and R. R. Rosa,
cond-mat/0010435
\bibitem{renyi} A. R\'{e}nyi, {\em Probability Theory}, North Holland,
Amsterdam (1970)
\bibitem{BS} C. Beck and F. Schl\"{o}gl, {\em Thermodynamics of
Chaotic Systems}, Camdridge University Press, Cambridge (1993)
\bibitem{abe} S. Abe, Phys. Lett. {\bf 271A}, 74 (2000)
\bibitem{web} http://tsallis.cat.cbpf.br/biblio.htm
\bibitem{sreeni} K.R. Sreenivasan and R.A. Antonia, Annu.
Rev. Fluid Mech. {\bf 29}, 435 (1997)
\bibitem{bod1} G.A. Voth, K.
Satyanarayan and E. Bodenschatz, Phys. Fluids {\bf 10}, 2268 (1998)
\bibitem{bod2} A. La Porta, G.A. Voth, A.M. Crawford, J. Alexander,
and E. Bodenschatz, Nature {\bf 409}, 1017 (2001)
\bibitem{vKa} N.G. van Kampen, {\em Stochastic Processes in Physics
and Chemistry}, North-Holland, Amsterdam (1981)
\bibitem{vedula} P. Vedula and P.K. Yeung, Phys. Fluids {\bf 11},
1208 (1999)
\bibitem{wilk} G. Wilk and Z. Wlodarczyk, Phys. Rev. Lett. {\bf 84}, 2770 (2000)
\bibitem{hast} N.A.J. Hastings and J.B. Peacock,
{\em Statistical Distributions}, Butterworth, London (1974)
\bibitem{BLS} C. Beck, G.S. Lewis and H.L. Swinney, Phys. Rev. {\bf 63E},
 035303(R) (2001) 
\end{thebibliography}
\end{document}